\documentclass[pra,twocolumn,showpacs,amsmath,amssymb,floatfix]{revtex4}
\usepackage{graphicx}
\usepackage{dcolumn}
\usepackage{bm}

\begin{document}

\title{Cavity-QED models of switches for attojoule-scale nanophotonic logic}
\author{Hideo Mabuchi} \email{hmabuchi@stanford.edu}
\affiliation{Edward L.\ Ginzton Laboratory, Stanford University, Stanford, California 94305, USA}

\date{\today}


\begin{abstract}
Quantum optical input-output models are described for a class of optical switches based on cavity quantum electrodynamics (cavity QED) with a single multilevel atom (or comparable bound system of charges) coupled simultaneously to several resonant field modes. A recent limit theorem for quantum stochastic differential equations is used to show that such models converge to a simple scattering matrix in a type of strong coupling limit that seems natural for nanophotonic systems. Numerical integration is used to show that the behavior of the pre-limit model approximates that of the simple scattering matrix in a realistic regime for the physical parameters, and that it is possible in the proposed cavity-QED configuration for low power optical signals to switch higher-power signals at attojoule energy scales.
\end{abstract}

\maketitle

\noindent It has long been appreciated that in cavity quantum electrodynamics (cavity QED) with strong coupling~\cite{Kimb98,Mabu02}, the transmission or reflection of an optical field coupled to the cavity mode can be controlled by the state of a single intra-cavity atom (or comparable bound system of charges,{\it e.g.}, as in solid-state systems)~\cite{Mabu96,Daya08,Khud09}. This simple insight leads naturally to various schemes for the implementation of optical switches in which a low-power beam is used somehow directly to manipulate the atomic internal state, and thus to determine whether a higher-power beam incident upon the cavity input coupling mirror is transmitted or reflected (for related proposals see~\cite{Berm06,Waks06,Chan07}). As cavity QED switches should be realizable in integrated nanophotonic platforms~\cite{Engl07,Barc09} and could potentially function down to the single photon level~\cite{Fara08} (although it is not clear whether robust circuit-level operation could be achieved with few-photon signals in the presence of finite propagation losses) this type of device could contribute substantially to the development of ultra-low power nanophotonic signal processing, for example in the context of on-chip nanophotonic interconnect~\cite{Mill00,Beau07,Mill09}.

In quantitative performance studies of cavity-QED switch designs it is important to consider fundamental models that include the effects of optical shot noise, spontaneous emission, and dipole fluctuations. Such models should be cascadable, and should clearly indicate the key parameters for optimizing switching performance. The purpose of this article is to describe quantum-optical input-output models for an elementary class of cavity QED switches, which can be cascaded easily using the series and concatenation products for quantum stochastic differential equations (QSDE's)~\cite{Goug09a}, and to show that these models converge to a very simple scattering-matrix description in a certain strong coupling limit that seems natural for nanophotonic systems. The limit model may provide a useful abstraction for circuit-level analysis and synthesis of photonic signal processing systems~\cite{Kerc09c,Jame07a,Nurd09a,Nurd09b,Mabu08b}, while numerical integration of the primary model can be used to characterize deviations from ideal switching behavior as well as the limits of low-power operation for finite values of the physical parameters.

\begin{figure}[tb!]
\includegraphics[width=0.48\textwidth]{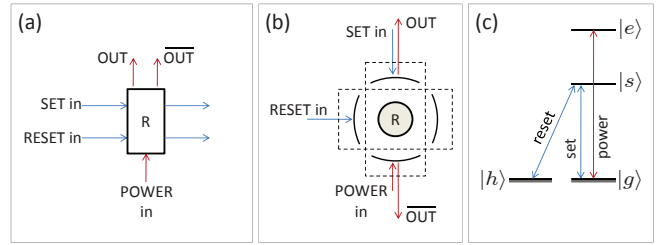}
\caption{\label{fig:sr-ff} Detail diagrams for a set-reset flip-flop switch (adapted from~\cite{Kerc09c}): (a) signal connection diagram, (b) physical diagram of optical inputs and outputs, and (c) required levels and couplings for the intra-cavity `atom.'}
\vspace{-0.1in}
\end{figure}

Fig.~\ref{fig:sr-ff} presents the basic structure of the cavity-QED switch we will consider, which was recently proposed in~\cite{Kerc09c}. The device functions essentially as a set-reset flip-flop switch. The intra-cavity atom, whose internal state determines the overall state of the switch, is assumed to have levels $\{\vert g\rangle,\vert h\rangle,\vert e\rangle,\vert s\rangle\}$ in the arrangement shown. States $\{\vert g\rangle,\vert s\rangle,\vert e\rangle\}$ are assumed to have the same angular momentum $m$ (eigenvalue of angular momentum along the $z$-axis) while $\vert h\rangle$ should have angular momentum $m-1$. Although we have drawn $\vert e\rangle$ and $\vert s\rangle$ at different energies, we note that this would not necessarily have to be the case if other selection rules (associated with a symmetry, rather than energy) could be used to prevent the POWER cavity mode from coupling to the $\vert g\rangle,\vert h\rangle\leftrightarrow\vert s\rangle$ transitions. In normal operation the atomic state should be either $\vert g\rangle$ or $\vert h\rangle$, as $\vert e\rangle$ is ideally never populated and $\vert s\rangle$ serves only to facilitate transitions between $\vert g\rangle$ and $\vert h\rangle$. We assume that the POWER input drives a cavity mode that couples only to the $\vert g\rangle\leftrightarrow\vert e\rangle$ transition, that the SET input drives a cavity mode that couples only to the $\vert g\rangle\leftrightarrow\vert s\rangle$ transition, and that the RESET input drives a cavity mode that couples only to the $\vert h\rangle\leftrightarrow\vert s\rangle$ transition. While the flexibility of, {\it e.g.}, photonic crystal resonator design could potentially provide the three required atom-field couplings via three modes in a single nanophotonic structure, we here propose a configuration of two Fabry-Perot-type cavities to illustrate concretely that the selection rules we have assumed are reasonable. Specifically, we assume that the cavities are sufficiently birefringent~\cite{Yang08} that the cavities support linearly-polarized resonant modes. In the POWER/SET cavity we utilize modes that are polarized along the $z$-axis (atomic quantization axis), so that these fields only induce $\pi$ transitions. In the RESET cavity we utilize a mode polarized along the $x$-axis, so that the associated field does not couple to $\pi$ transitions but can couple the $\sigma_+$ transition $\vert h\rangle\leftrightarrow\vert s\rangle$. We assume that the cavities are symmetric (as opposed to single-sided).

We specify our primary switch model via the $({\bf S},{\bf L},H)$ coefficients~\cite{Goug09a} for a (right-sided) QSDE with ${\bf S}=I$, the components of the vector ${\bf L}$ given by
\begin{eqnarray}
L_1&=&L_2=k_1\sqrt{\kappa_p}a,\quad L_3=L_4=k_2\sqrt{\kappa_s}b,\nonumber\\
L_5&=&L_6=k_2\sqrt{\kappa_r}c,\quad L_7=\sqrt{\Gamma}\sigma_{gs},\quad L_8=\sqrt{\Gamma}\sigma_{hs},\nonumber\\
L_9&=&\sqrt{\Gamma}\sigma_{ge},\quad L_{10}=\sqrt{\Gamma}\sigma_{he},\label{eq:SLprimary}
\end{eqnarray}
and
\begin{eqnarray}
H&=&ik_1^2g_p(a^\dag\sigma_{ge}-a\sigma^\dag_{ge})+ik_2g_s(b^\dag\sigma_{gs}-b\sigma^\dag_{gs})\nonumber\\
&&+ik_2g_r(c^\dag\sigma_{hs}-c\sigma^\dag_{hs}),\label{eq:Hprimary}
\end{eqnarray}
where $a$, $b$ and $c$ are annihilation operators for the POWER, SET and RESET cavity modes, $\sigma_{ge}\equiv\vert g\rangle\langle e\vert$, $\sigma_{he}\equiv\vert h\rangle\langle e\vert$, $\sigma_{gs}\equiv\vert g\rangle\langle s\vert$, and $\sigma_{hs}\equiv\vert h\rangle\langle s\vert$. The parameter $\Gamma$ is an atomic spontaneous emission rate, $\kappa_{p,s,r}$ and $g_{p,s,r}$ are the field decay rates and atomic coupling strengths (vacuum Rabi frequencies) of the POWER, SET and RESET cavity modes, and $k_{1,2}$ are dimensionless scaling parameters that we will use to derive the scattering-matrix limit. Note that the rates of the four atomic spontaneous emission processes could vary with minor impact on switch performance. With the above ordering of the components of ${\bf L}$ we specify QSDE modes $1$ and $2$ as the POWER input and output, modes $3$ and $4$ as the SET input and output, and modes $5$ and $6$ as the RESET input and output.

The operating principle of the switch is as follows. With the atom in the $\vert h\rangle$ state, the POWER cavity mode does not couple to the atom at all and therefore the POWER input field is routed to the OUT output (transmitted through the cavity). With the atom in the $\vert g\rangle$ state, however, the POWER cavity mode experiences a large vacuum Rabi splitting and the POWER input is therefore routed to the $\overline{\rm OUT}$ output (reflected from the cavity). When the atom is in the $\vert g\rangle$ state, injection of photons into the SET cavity mode induces Rabi oscillation on the $\vert g\rangle\leftrightarrow\vert s\rangle$ transition that is promptly terminated by decay into the atomic $\vert h\rangle$ state via cavity-enhanced (via coupling to the RESET mode vacuum) spontaneous emission. When the atom is in the $\vert h\rangle$ state, decay into the $\vert g\rangle$ state can likewise be induced by injection of photons into the RESET cavity mode. We assume that under normal operation photons are never injected into both the SET and RESET cavity modes simultaneously, as this would induce a `race' condition which should not normally occur in a classical logic circuit. Note that when no photons are injected into either the SET or RESET cavity mode, the atomic state should ideally remain constant (`hold' condition), although again this should not be necessary in conventional circuits.

\begin{figure}[tb!]
\includegraphics[width=0.48\textwidth]{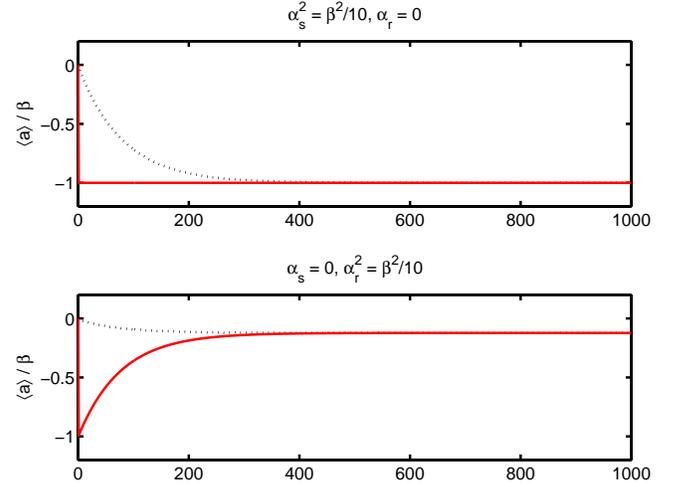}
\caption{\label{fig:num} Numerical integrations of the pre-limit master equation, displaying the intra-cavity field amplitude (normalized by $\beta$) as a function of time. In each subplot, the black (broken) curve is for initial state $\vert g\rangle$ and the red (solid) curve is for initial state $\vert h\rangle$. See text for parameters.}
\vspace{-0.1in}
\end{figure}
In order to prove that this primary model should behave according to these principles in a regime of strong coupling we invoke the QSDE limit theorem of~\cite{Bout08}, first taking the scaling parameter $k_1\rightarrow\infty$ and subsequently taking $k_2\rightarrow\infty$. For the first limit we satisfy the structural requirements of~\cite{Bout08} by choosing the limit state-space $H_0={\rm span}\{\vert g\,0_a n_b n_c\rangle,\vert h\,0_a n_b n_c\rangle,\vert s\,0_a n_b n_c\rangle\}$ (where the photon numbers $n_{a,b,c}$ indicate corresponding Fock states of the cavity modes), and for the second limit we choose $H_0={\rm span}\{\vert g\,0_a 0_b 0_c\rangle,\vert h\,0_a 0_b 0_c\rangle,\vert s\,0_a 0_b 0_c\rangle\}$. In the resulting model---which is actually still an intermediate model from which we will subsequently derive an even simpler scattering matrix---the $\vert e\rangle$ excited state and all three cavity modes have been adiabatically eliminated. As a result we can drop modes $9$ and $10$ from the QSDE model, as they were used only to provide spontaneous emission from $\vert e\rangle$. The remaining coefficients of the `intermediate' QSDE are
\begin{eqnarray}
S_{11}&=&S_{22}=\Pi_g,\quad S_{12}=S_{21}=-\Pi_{hs},\nonumber\\
S_{34}&=&S_{43}=S_{56}=S_{65}=S_{77}=S_{88}=1,\nonumber\\
L_3&=&L_4=\sqrt{\gamma}\sigma_{gs},\quad L_5=L_6=\sqrt{\gamma}\sigma_{hs},\nonumber\\
L_7&=&\sqrt{\Gamma}\sigma_{gs},\quad L_8=\sqrt{\Gamma}\sigma_{hs},\label{eq:mid}
\end{eqnarray}
with $H$ and all remaining components of ${\bf S}$ and ${\bf L}$ equal to zero. Here $\Pi_g\equiv \vert g\rangle\langle g\vert$, $\Pi_{hs}\equiv \vert h\rangle\langle h\vert + \vert s\rangle\langle s\vert$, and we have defined a new parameter $\gamma\equiv g^2/\kappa$ (which we will later limit $\rightarrow\infty$) assuming $g_s=g_r\equiv g$ and $\kappa_s=\kappa_r\equiv\kappa$.

We note that in the context of a circuit analysis the series product for QSDE's~\cite{Goug09a} can be used to connect the inputs and outputs of our switch model to those of other components. For illustrative purposes here we simply apply coherent-state inputs of fixed amplitude $\beta$, $\alpha_s$ and $\alpha_r$ to the POWER, SET and RESET inputs (note that in this type of QSDE model, coherent amplitudes are normalized such that their square-magnitudes correspond to photon flux per unit time). This results in the simple modifications
${\bf L}\mapsto {\bf L} + {\bf S}{\bf d}$ and $H\mapsto H + {\rm Im}\{{\bf L}^\dag{\bf S}{\bf d}\}$, where ${\bf d}$ is a column vector of displacement amplitudes ($d_1=\beta$, $d_3=\alpha_s$, $d_5=\alpha_r$ and all other components zero). The master equation corresponding to a QSDE model is in general
\begin{equation}
\dot{\rho}_t=-i[H,\rho_t] + \sum_i\left( L_i\rho_tL_i^* - \frac{1}{2}\{L_i^*L_i,\rho_t\}\right),\label{eq:meqgeneric}
\end{equation}
which now takes the form
\begin{eqnarray}
\dot{\rho}_t&=&-i[H',\rho_t] + (\Gamma+2\gamma)\left( \sigma_{gs}\rho_t\sigma_{gs}^\dag - \frac{1}{2}\{\Pi_s,\rho_t\}\right)\nonumber\\
&&+(\Gamma+2\gamma)\left( \sigma_{hs}\rho_t\sigma_{hs}^\dag - \frac{1}{2}\{\Pi_s,\rho_t\}\right)\nonumber\\
&&+\vert\beta\vert^2\left(\Pi_g\rho_t\Pi_g + \Pi_{hs}\rho_t\Pi_{hs} - \rho_t\right),\label{eq:meqmid}
\end{eqnarray}
with
\begin{equation}
H'=i\sqrt{\gamma}(\alpha_s\sigma_{gs}^\dag - \alpha_s^*\sigma_{gs})+i\sqrt{\gamma}(\alpha_r\sigma_{hs}^\dag - \alpha_r^*\sigma_{hs}).\label{eq:Hmid}
\end{equation}
Note that this $H'$ includes terms from originating from the Lindblad part of Eq.~(\ref{eq:meqgeneric}). It is easy to see that the terms in $H'$ induce the atomic Rabi oscillations mentioned in our description of the switch operating principles above, while the Lindblad terms provide the required cavity-enhanced spontaneous emission (with rate $\Gamma+2\gamma$) from  $\vert s\rangle$ to $\vert g\rangle,\vert h\rangle$ as well as dephasing (at rate $\vert\beta\vert^2$) of $\vert g\rangle$ relative to the subspace spanned by $\vert h\rangle$ and $\vert s\rangle$. The dephasing occurs since an outside observer can easily distinguish which of these subspaces the atomic state is in, simply by measuring the optical power in the OUT and $\overline{\rm OUT}$ outputs, and is generally consistent with the desired switch dynamics. It would appear, however, that for $\vert\beta\vert\gg\vert\alpha_r\vert$ there should be some `Zeno-like' suppression of the reset action which can limit the degree of signal regeneration $\vert\beta\vert/\vert\alpha_{r,s}\vert$ that can be accomplished by this type of switch without loss of switching speed.
When $\alpha_s=\alpha_r=0$ (hold condition) we have relay equilibrium states $\vert g\rangle$ and $\vert h\rangle$. Looking at the upper $2\times 2$ block of ${\bf S}$ we find that it becomes the identity for $\vert g\rangle$, indicating perfect reflection of the POWER input beam into $\overline{\rm OUT}$, while for $\vert h\rangle$ it indicates transmission of the POWER input into OUT. With $\alpha_s=0,\alpha_r\ne 0$ (RESET condition) only $\vert g\rangle$ is an equilibrium state while with $\alpha_s\ne 0,\alpha_r=0$ (SET condition) only $\vert h\rangle$ is an equilibrium state, so we find clear agreement between the structure of this intermediate model and the switch operating principle described above.

We now take a final limit $\gamma\rightarrow\infty$ to obtain a very simple scattering matrix model of an idealized cavity-QED switch. Returning to the un-driven intermediate QSDE model of Eq.~(\ref{eq:mid}) and using once again the limit theorem of~\cite{Bout08} with $H_0={\rm span}\{\vert g\,0_a 0_b 0_c\rangle,\vert h\,0_a 0_b 0_c\rangle\}$, we obtain the limit QSDE coefficients
\begin{eqnarray}
S_{11}&=&S_{22}=\Pi_g,\quad S_{12}=S_{21}=-\Pi_h,\nonumber\\
S_{33}&=&S_{44}=-\Pi_g/2,\quad S_{34}=S_{43}=1-\Pi_g/2,\nonumber\\
S_{55}&=&S_{66}=-\Pi_h/2,\quad S_{56}=S_{65}=1-\Pi_h/2,\nonumber\\
S_{35}&=&S_{36}=S_{45}=S_{46}=-\sigma_{gh}/2,\nonumber\\
S_{53}&=&S_{54}=S_{63}=S_{64}=-\sigma_{hg}/2,\label{eq:lim}
\end{eqnarray}
where $\Pi_h\equiv\vert h\rangle\langle h\vert$, $\sigma_{hg}\equiv\vert h\rangle\langle g\vert$, and $\sigma_{gh}\equiv\vert g\rangle\langle h\vert$. All remaining components of ${\bf S}$ as well as both ${\bf L}$ and $H$ are zero. If we once again consider displacements of input modes 1, 3 and 5 with coherent amplitudes $\beta$, $\alpha_s$ and $\alpha_r$ respectively, the resulting master equation for $\rho_t$ can be written in terms of independent matrix elements in the $\{\vert g\rangle,\vert h\rangle\}$ basis,
\begin{eqnarray}
\dot{\rho}_{gg}&=&\frac{1}{2}\left(-\vert\alpha_s\vert^2\rho_{gg}+\vert\alpha_r\vert^2\rho_{hh}\right),\nonumber\\
\dot{\rho}_{hg}&=&-\frac{1}{2}\alpha_s\alpha_r^* - \rho_{hg}\left(\vert\beta\vert^2+\vert\alpha_s\vert^2/2+\vert\alpha_r\vert^2/2\right).\label{eq:limme}
\end{eqnarray}
We thus find that the rates of SET/RESET action are given in this limit by $\vert\alpha_{s,r}\vert^2/2$. At equilibrium we find
\begin{eqnarray}
\rho_{gg}&\rightarrow&\vert\alpha_r\vert^2/\left(\vert\alpha_s\vert^2+\vert\alpha_r\vert^2\right),\nonumber\\
\rho_{hg}&\rightarrow&-\frac{\alpha_s\alpha_r^*}{2\vert\beta\vert^2+\vert\alpha_s\vert^2+\vert\alpha_r\vert^2},
\end{eqnarray}
which shows that in fact sensible operation can be expected even with $\alpha_s\alpha_r\ne 0$ (race condition).

As mentioned in the introduction, it is straightforward to perform numerical integration of the primary (pre-limit) model given in Eqs.~(\ref{eq:SLprimary}) and (\ref{eq:Hprimary}) to characterize the switch performance for any desired set of finite values of the physical parameters. In Fig.~\ref{fig:num} we display the results of such an integration, performed using the Quantum Optics Toolbox for Matlab~\cite{Tan99}, for parameter values $\Gamma=0.3$, $g_p=50$, $g_{s,r}=10$, $\kappa_{p,s,r}=50$, $\beta=0.5$, and $\alpha_{s,r}$ as indicated in each subplot (with $k_{1,2}$ fixed at unity). Assuming that $g$ values generally can be decreased and $\kappa$ values increased via resonator design optimization, this hierarchy of rates is consistent with the parameters $(g,\kappa,\Gamma)/2\pi=(16,16,0.1)$ GHz of current work with GaAs nanophotonic resonators and InAs quantum dots~\cite{Fara08} (although quantum dots would not seem to have the multi-level structure required for our scheme), and with projected numbers $(g,\kappa,\Gamma)/2\pi=(2.25,0.16,0.013)$ GHz for GaP nanophotonic resonators and diamond-NV centers~\cite{Barc09}.

\begin{figure}[tb!]
\includegraphics[width=0.38\textwidth]{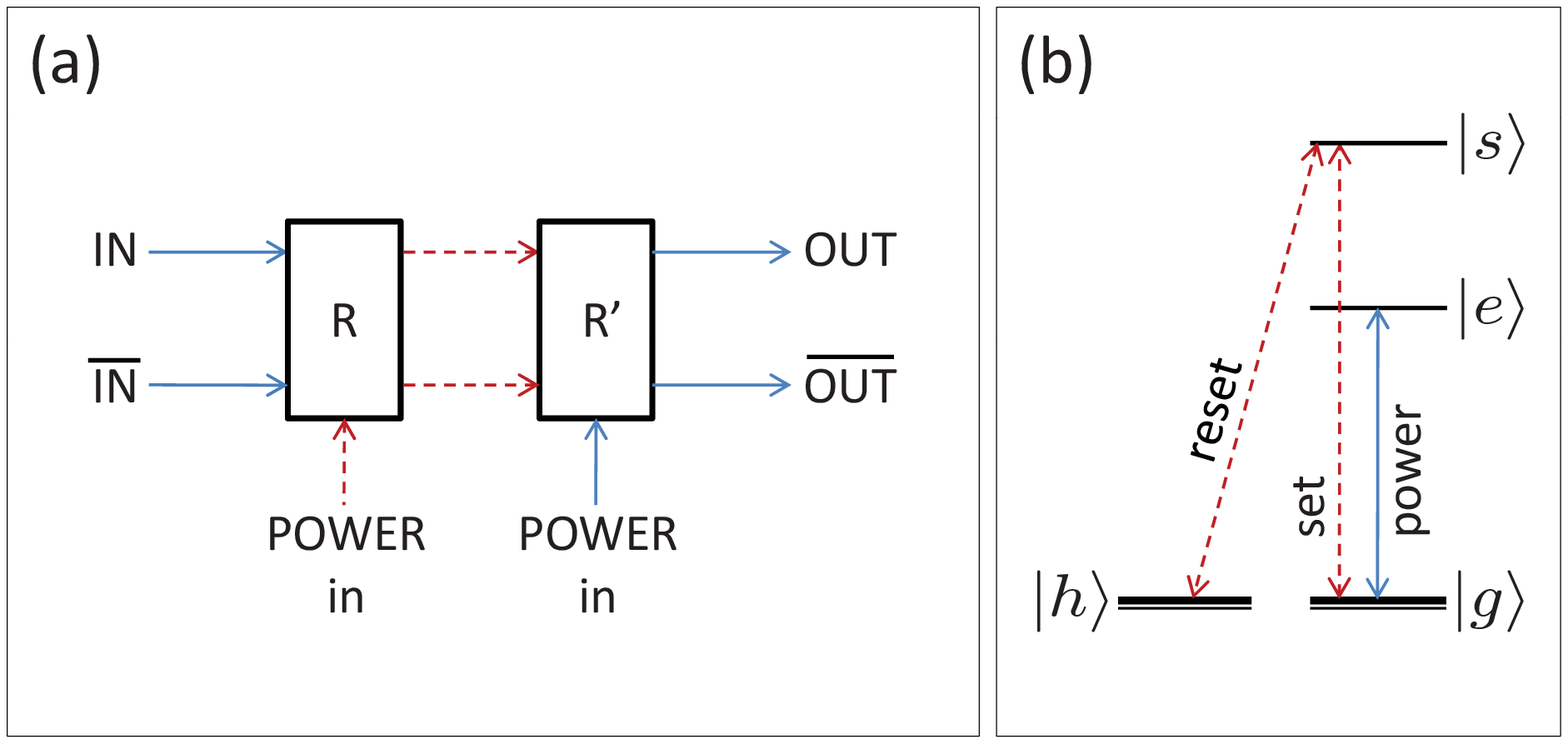}
\caption{\label{fig:updown} Configuration of two elementary switches to match input and output wavelengths, in case this cannot be done at the individual switch level: (a) port connections, and (b) modified coupling diagram in switch $R'$ (note that the atomic level structure is the same in $R$ and $R'$).}
\end{figure}

\begin{figure}[t!]
\includegraphics[width=0.48\textwidth]{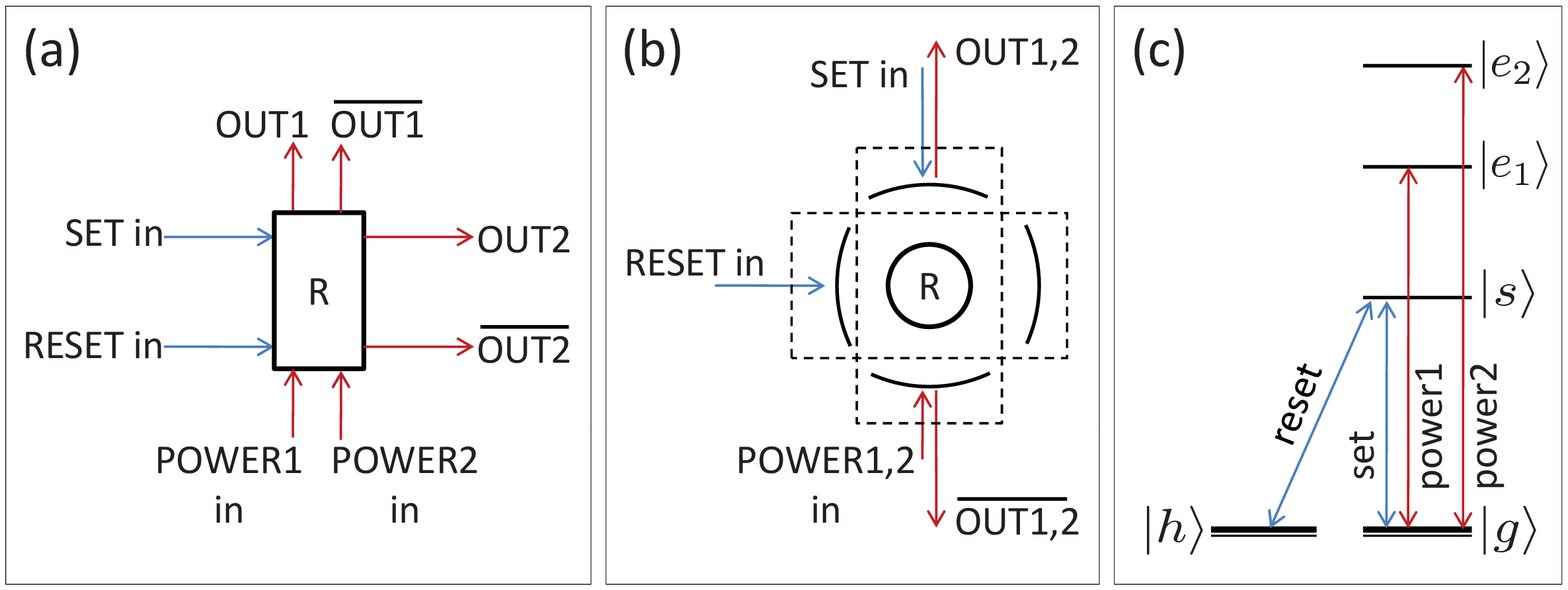}
\caption{\label{fig:sr-dpdt} Detail diagrams for a set-reset double-pole double-throw (DPDT) switch: (a) optical inputs and outputs, (b) required level diagram and couplings for the intra-cavity `atom.'}
\vspace{-0.1in}
\end{figure}
From the usual cavity QED input-output rule for two-sided resonators~\cite{Wall95,Goug09a} we have that the coherent amplitude of the reflected $\overline{\rm OUT}$ mode will be $\sqrt{\kappa_p}\, a+\beta$ while that of the transmitted OUT mode will be $\sqrt{\kappa_p}\, a$. The results in Fig.~\ref{fig:num} thus clearly show that when the SET input is active the relay is driven to the $\vert h\rangle$ state in which OUT is high, whereas when the RESET input is active it is driven to the $\vert g\rangle$ state in which $\overline{\rm OUT}$ is high. In the latter condition a finite contrast (power) ratio $\overline{\rm OUT}$/OUT $\approx 66$ results from the finite ratios $g_{p,s,r}/(\kappa_{p,s,r},\Gamma)$, but assuming this is acceptable the results show that switching is achieved with a power `gain' of $\vert\beta\vert^2/\vert\alpha_{s,r}\vert^2=10$. We furthermore see that the switching occurs within a timescale $\tau\sim 300$ and is thus effected by $\tau\vert\alpha_{s,r}\vert^2\sim 10$ photons, corresponding to a switching energy in the aJ range assuming visible or near-IR wavelengths.

While we have proposed a concrete model in which the POWER beam differs in wavelength from the SET/RESET beams (in order to utilize an atomic model with simple selection rules), it should be possible to create circuits from such components in which many sequential logic operations are performed with only two distinct signal wavelengths. To illustrate this point, we sketch in Fig.~\ref{fig:updown}a a cascade of two switches that achieves signal power regeneration without overall change in wavelength. The first switch $R$ is of the type we have considered above while the second switch $R'$ is an analogous device constructed with the modified coupling diagram shown in Fig.~\ref{fig:updown}b. We likewise note that our basic design can be extended straightforwardly to design more complex switches. In Fig.~\ref{fig:sr-dpdt} for example we illustrate the basic principle of a device in which a single pair of SET/RESET input beams simultaneously switch two POWER beams with different wavelengths~\cite{Kerc09c}; the POWER1 and POWER2 beams are assumed to probe distinct longitudinal modes of the cavity.

This work was supported by DARPA-MTO/ARO (W911NF-09-1-0045) and IARPA/ARO (W911NF-08-1-0491). The author thanks Hendra Nurdin and Joseph Kerckhoff for useful discussions.

\end{document}